\documentclass[superscriptaddress,
twocolumn, 
amssymb, 
prab
]{revtex4-2} % single column

\usepackage{amsmath}
\usepackage[utf8]{inputenc}
\usepackage{url}
\usepackage{hyperref}% add hypertext capabilities
\hypersetup{
    colorlinks=true,
    linkcolor=blue,
    filecolor=blue,      
    urlcolor=blue,
}
\usepackage{graphicx}
\usepackage{epstopdf}
\usepackage{caption}
\usepackage{subcaption}
\usepackage{color}
\usepackage{dcolumn}% Align table columns on decimal point
\usepackage{bm}% bold math\
\usepackage{upgreek}
\usepackage{comment}
\usepackage{ulem} % https://jansoehlke.com/2010/06/strikethrough-in-latex/
\usepackage[english]{babel}
\newcommand{\avg}[1]{\langle#1\rangle}

\begin{document}

\title[Electron beam shaping via laser heater temporal shaping]{Electron beam shaping via laser heater temporal shaping}

\author{D. Cesar}
\email{dcesar@slac.stanford.edu}
\affiliation{SLAC National Accelerator Laboratory, Menlo Park, CA}

\author{A. Anakru}
\affiliation{Cornell University, Ithaca, NY, USA}

\author{S. Carbajo}
\affiliation{SLAC National Accelerator Laboratory, Menlo Park, CA}

\author{J. Duris}
\affiliation{SLAC National Accelerator Laboratory, Menlo Park, CA}

\author{P. Franz}
\affiliation{SLAC National Accelerator Laboratory, Menlo Park, CA}

\author{S. Li}
\affiliation{SLAC National Accelerator Laboratory, Menlo Park, CA}

\author{N. Sudar}
\affiliation{SLAC National Accelerator Laboratory, Menlo Park, CA}

\author{Z. Zhang}
\affiliation{SLAC National Accelerator Laboratory, Menlo Park, CA}

\author{A. Marinelli}
\email{marinelli@slac.stanford.edu}
\affiliation{SLAC National Accelerator Laboratory, Menlo Park, CA}

\date{\today}

\begin{abstract}
Active longitudinal beam optics can help FEL facilities achieve cutting edge performance by optimizing the beam to: produce multi-color pulses, suppress caustics, or support attosecond lasing. As the next generation of superconducting accelerators comes online, there is a need to find new elements which can both operate at high beam power and which offer multiplexing capabilities at Mhz repetition rate. Laser heater shaping promises to satisfy both criteria by imparting a programmable slice-energy spread on a shot-by-shot basis. We use a simple kinetic analysis to show how control of the slice energy spread translates into control of the bunch current profile, and then we present a collection of start-to-end simulations at LCLS-II in order to illustrate the technique.
\end{abstract}

% \pacs{}
% 41.60.Cr - Free-electron lasers
% 42.55.Vc - X- and γ-ray lasers

\maketitle
%===========================================================================================
%===========================================================================================
%===========================================================================================
%===========================================================================================

\section{Introduction}
Facility scale electron accelerators regularly produce high current, high-brightness electron bunches by accelerating and compressing the bunch from an RF photo-injector. During bunch compression, collective forces and nonlinear optics passively act on the bunch and unavoidably alter the current profile, such that many applications can benefit by actively re-shaping the intra-bunch current profile. For example, in the context of x-ray free-electron lasers (XFEL), modulation of the current profile can lead to: optimized beam brightness, suppressed horn formation \cite{ding_beam_2016,prat_compact_2020}, two-color operation \cite{roussel_multicolor_2015}, and attosecond x-ray pulses \cite{duris_tunable_2020, zhang_experimental_2020, malyzhenkov_single-and_2020, huang_generating_2017}; while in plasma-wakefield accelerators, shaping of the current profile can lead to maximization of the energy transfer to the witness bunch in high-transformer ratio experiments \cite{loisch_observation_2018,andonian_generation_2017}.

Existing beam-shaping techniques are often reliant on support from solid-target based collimators \cite{ding_beam_2016} and emittance spoilers \cite{marinelli_experimental_2017,emma_femtosecond_2004}. But as the next generation of high average power accelerators \cite{emma_linear_2014,decking_commissioning_2017} come on-line, these must be replaced by new techniques which are compatible with hundreds of kilowatts of electron beam power. Laser heater shaping is an attractive alternative because it provides a highly flexible platform for bunch shaping at MHz repetition rate. 

 The laser heater is a laser modulator located in the middle of a chicane \cite{huang_measurements_2010}, such that dispersion destroys the microbunching structure and leaves behind a symmetric energy spread. Conventionally, laser heaters are used to suppress the collective instabilities that develop during acceleration and bunch compression (namely the microbunching instability, \cite{huang_suppression_2004,huang_measurements_2010}).  To accomplish this, the laser duration is much longer than the beam duration such that the heating is applied uniformly along the electron beam. If instead the laser heater intensity profile is modulated as a function of time, then we will create a time-dependent energy spread which can be used to shape the bunch current profile.
 
Such temporal laser shaping was initially proposed to control the temporal properties of XFELs \cite{marinelli_optical_2016} and multicolor generation in a seeded FEL \cite{roussel_multicolor_2015}. More recently, a highly periodic laser heater modulation has been used to manipulate the microbunching instability \cite{brynes_microbunching_2020}. Laser heater shaping has also been proposed to generate a train of current spikes in order to seed a high-power THz wiggler \cite{zhang_generation_2017}. Here we generalize these methods to the generation of complex current profiles by an arbitrarily shaped laser heater.

High resolution adaptive temporal shaping of broadband femtosecond laser pulses can be achieved via spectral phase and amplitude shaping or programmable synthesis. For instance, Volume Bragg gratings\cite{lumeau_tunable_2006} can be readily used to create single and multiple notches for emittance spoiling or beam slicing, thanks in part to their high diffraction efficiency, high damage threshold, and temperature and optical stability. In the case of more sophisticated or adaptable temporal shapes, spectral amplitude and phase modulators\cite{hornbeck_spatial_1991,tournois_acousto-optic_1997} can be used, opening the door to machine-learning-based control, albeit at damage-limited low-to-moderate peak and average power levels\cite{carbajo_power_2018} or with a substantially reduced actuation range after amplification of spectrally shaped pulses . Recent demonstrations in the synthesis of 4D programmable light bullets with adaptable spatio-temporal shapes\cite{lin_reconfigurable_2020,lemons_integrated_2020} promise to overcome some of these power limitations in the near future.

The article is organized as follows: first we introduce a general theoretical model that describes the evolution of the current profile for an arbitrary laser heater temporal profile going through a single bunch compressor. We then provide a series of numerical case studies in which a temporally shaped laser heater provides useful control of the current profile at an XFEL.

\begin{figure}[ht]
    \centering
    \includegraphics[width=0.45\textwidth]{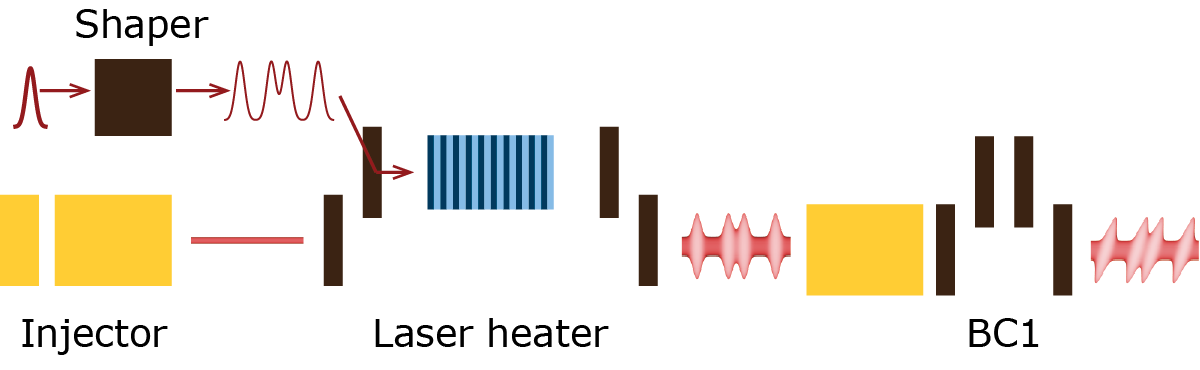}
    \caption{Cartoon of laser heater shaping at an FEL facility. A high repetition rate laser can be used to program complicated patterns into the slice energy spread. After bunch compression (BC1) the energy spread is converted to a current modulation.}
    \label{fig:cartoon_intro}
\end{figure}
%===========================================================================================
%===========================================================================================
%===========================================================================================
%===========================================================================================
\section{Analytical model}

%Introduce theoretical model
Active manipulation of the beam's slice energy spread is an indirect method for shaping its current profile. It works by modulating the beam's longitudinal phase space, such that, immediately following the laser heater, the  current is unperturbed, but after bunch compression the energy spread is converted into a current modulation. Even small current modulations can seed further growth downstream via the micro-bunching instability mechanism. To understand the extent to which the laser heater can be used to draw structure into the beam's longitudinal phase space, we start with a simple kinetic analysis describing the compression of a beam with slice-dependent energy spread.  We then apply the results of this analysis to study a series of applications: the generation of isolated current spikes by a single Gaussian laser heater, a quasi-periodic current profile made by stacking Gaussian pulses, and a highly-periodic current profile made from a laser comb.

%Assumptions. Energy modulation
In order to isolate the effect of the laser heater, we consider a simplified model in which the longitudinal phase space is infinite in $s$ and uncoupled from $\eta=(\gamma-\gamma_0)/\gamma_0$. The initial current profile has a flat-top current of $I_0$. The laser heater intensity is assumed to be uniform in the transverse plane such that the slice-wise energy modulation of the beam will follow an arcsin distribution in the $\eta$ variable---for more complicated transverse profiles \cite{liebster_laguerre-gaussian_2018,tang_laguerre-gaussian_2020} one must make suitable adjustments. Then, the phase space distribution of the electron bunch after the laser heater is given by

$$f_0(s,\eta) =  \frac{1}{\pi}\frac{1}{\sqrt{A(s)^2-\eta^2}}$$

where $A(s)$ describes the slice energy spread of the beam as a function of $s$. Neglecting slippage and dispersion in the laser heater, $A(s)$ is directly proportional to the square root of the laser power profile:

$$A(s)=\sqrt{2 Z_0 I_l(s)}\frac{q_e K L_u JJ}{2 \gamma_0 \gamma m c^2}$$

where $I_l(s)\approx 2P(s)/\pi w_0^2$ is the on-axis laser intensity, $\gamma mc^2$ is the electron energy, $K$ is the undulator parameter, $L_u$ is the undulator magnetic length, $\gamma_0$ is the reference beam energy and $JJ$ is given by:
\begin{equation}
    JJ=J_0\left( \frac{K^2}{4+2K^2} \right)-J_1\left( \frac{K^2}{4+2K^2} \right)
\end{equation}
where $J_n$ are the Bessel functions of the first kind.
%Bunch compression

A longitudinally dispersive device, such as a bunch compressor, can be described by applying the following change of coordinates $s=s_{0}+R_{56}\eta_{0}$, where a subscript of zero denotes a coordinate before transport. Following Liouville's theorem, the phase-space distribution after the dispersive section is given by:

$$f(s,\eta) =f_0(s_0,\eta_0)=f_0(s-R_{56}\eta,\eta)$$

Then the current will be given by computing the marginal distribution in $s$, 

$$I(s) = I_0\int f_0(s-R_{56}\eta,\eta)d\eta $$

%Fourier, bunching factor
It will be convenient to work with the Fourier transform of the current, given by

%There's a factor of root(2*pi) in a typical FT definition, but it's nice not to have to write it. We must remember that it exists in the IFT then!
\begin{align}
    \mathcal{F}[I](k) &= I_0\int e^{-ik(s+R_{56}\eta)}f_0(s,\eta)dsd\eta \nonumber\\ 
    &=I_0\int   e^{-iks}\biggr(\int e^{-ikR_{56}\eta}f_0(s,\eta)d\eta \biggr)ds \nonumber
\end{align}

%Series expansion
The inner integral is the Fourier transform of an arcsin distribution, which can be formally evaluated as $J_0(kR_{56}A(s))$, however to proceed analytically it is helpful to expand the exponential in a power series about $R_{56}=0$. Then we will have a series involving moments of the modulation amplitude:  $\avg{A^n(s)} = \int \eta^n f_0(s,\eta)d\eta$. For any symmetrical distribution, the odd moments vanish by parity, and for arcsin the remaining sum can be written as:
$$\mathcal{F}[I](k)= I_0\int ds e^{-iks}\sum_{n=0}^\infty \left(\frac{i R_{56}^{n}}{2^{n} n!}\right)^2 (ik)^{2n} A^{2n}(s) $$

%% Inversion
Finally we can invert the Fourier transform term by term in the series. 
\begin{equation}
\label{eq:simple_current_modulation}
    I(s) = I_0 \sum_{n=0}^\infty \left(\frac{R_{56}^{n}}{2^{n} n!}\right)^2 \frac{\partial^{2n}}{\partial s^{2n}} A^{2n}(s)
\end{equation}

% Interpretation of series result
Equation \ref{eq:simple_current_modulation} gives our most intuitive picture of how laser heater shaping works. The $n=0$ term gives the unmodified current, and the $n=1$ term gives the leading order modulation. The amplitude of the $n=1$ term immediately shows us that, to leading order, the bunching is proportional to the laser power ($A_0^2$), rather than the electric field ($A_0$). This reflects the indirect nature of bunch shaping via laser heater: linear contributions are washed out by the symmetry of the energy spread. The remaining first order current profile is shaped like the second derivative of the laser power, such that a local maxima of the energy spread is transformed into a current minima, flanked by two smaller current peaks. Higher order terms sharpen the peaks and saturate the valley (which, after all, cannot go below zero current). Total charge is preserved at each order.

%Generalizations to more complicated cases
We can generalize Eq.\ref{eq:simple_current_modulation} to the more realistic case in which the initial distribution has a Gaussian slice energy spread $\sigma_\eta$ convolved with the arcsin distribution. We also include a linear chirp $h$ which will compress the entire beam by a factor of $C=(1+hR_{56})^{-1}$. The result is: 
\begin{align}\label{eq:general_modulation}
I(s) = C I_0 \sum_{n=0}^\infty \biggr[ & \left(\frac{R_{56}^{n}}{2^{n} n!}\right)^2 \times \nonumber \\
&   \frac{\partial^{2n}}{\partial s^{2n}}\left(A^{2n}(Cs) \circledast \frac{e^{-\frac{s^2}{2\left(R_{56}\sigma_\eta\right)^2}}}{\sqrt{2\pi}R_{56}\sigma_\eta}\right) \biggr]
\end{align}
%interpretation of damping.
The chirp gives an overall scaling, while the incoherent energy spread is only important if $R_{56}\sigma_\eta$ is comparable to the length scale of the laser heater modulation. The energy spread becomes increasingly important when compared to higher order terms in the series, especially as we increase $R_{56}$ to approach full compression. For a modulation of magnitude $A_0$ occurring over a characteristic time scale $\sigma_t$, we can expect that the dispersion required for full compression would be $R_{56}\approx c\sigma_t/ C A_0$. For these parameters, Eq.\,\ref{eq:simple_current_modulation} suggests the incoherent energy spread will be unimportant if $\sigma_\eta\lessapprox A_0$.

%Limits of the formula
However, it is precisely as we approach full compression that the series solution presented here begins to require a large number of terms to reach convergence. This happens because as the phase space folds back on itself and begins to filament, the current spikes stop growing and begin changing shape. Thus the limit $\sigma_\eta\lessapprox A_0$ applies only for the first term in the series, and near full compression we must always take $\sigma_\eta$ into account to determine the shape of the current spike. In many cases, this is actually a useful way to operate the heater, since the overheated, over-compressed electrons can be used to prevent undesired current spikes from forming at the edges of the beam. Nonetheless, the series solution gives a clear picture of how coherent microbunching arises from a time-dependent energy spread, and it shows how the laser heater parameters may be used to control the longitudinal phase space of the bunch.

%===========================================================================================
%===========================================================================================
%===========================================================================================
%===========================================================================================

\subsection{Gaussian}
\label{sec:analytical_Gaussian}
In this section we consider specialization to a single pulse with a Gaussian power profile:

$$A(s) = A_0e^{-\frac{1}{4}\frac{s^2}{c^2\sigma_t^2}}$$

In this case, the convolution in Eq.\,\ref{eq:general_modulation} can be easily evaluated. Identifying $\tilde{\sigma}_n^2=c^2\sigma_t^2/nC^2 + R_{56}^2\sigma_\eta^2$, we can write:

\begin{align}
\label{eq:Theory_Gaussian}
\begin{split}
I(s)=C I_0 \sum_{n=0}^\infty \biggr[&\left(\frac{R_{56} A_0}{2 \tilde{\sigma}_n} \right)^{2n} \frac{c \sigma_t}{C \sqrt{n} \tilde {\sigma}_n} \frac{(2n)! (-1)^n}{2^{n} (n!)^3} \times  \\
&\left(\frac{\tilde{\sigma}_n^{2n}2^{n}(n!)}{ (2n!) (-1)^n}\right)\frac{\partial^{2n}}{\partial s^{2n}}\left(e^{-\frac{1}{2}\frac{s^2}{\tilde{\sigma}_n^2}}\right) \biggr]
\end{split}
\end{align}
where the second line has been written such that it evaluates to 1 at $s=0$ for all $n$, and thus the sum of the top line alone gives the current at the $t=0$ minimum.

\begin{figure}[ht]
    \centering
    \includegraphics[width=0.45\textwidth]{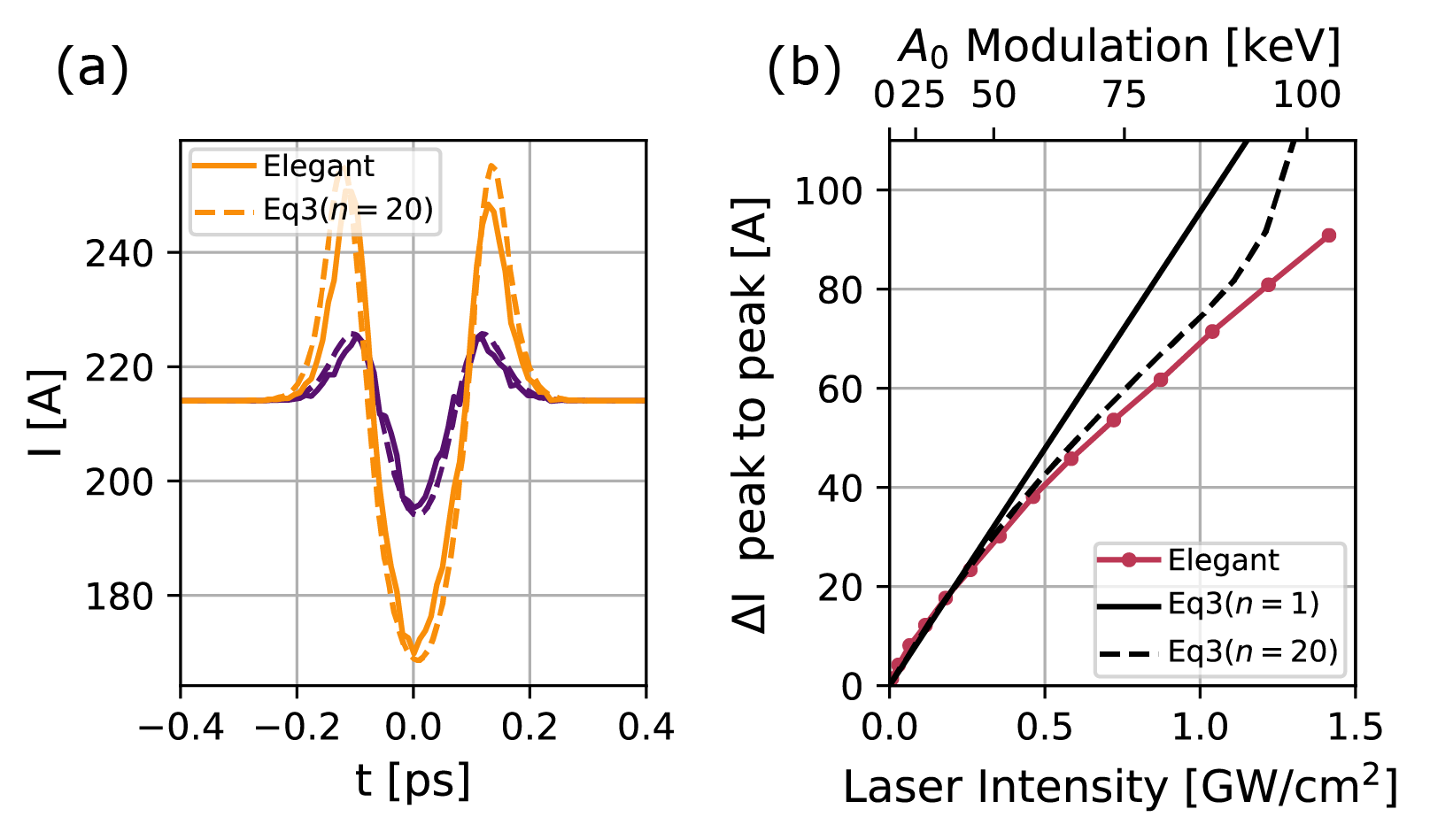}
    \caption{Comparison of analytical model with Elegant simulation of a laser heater plus bunch compressor. (a) Current profiles for small and large laser intensity. At large intensity multiple terms are required for the series solution to converge. (b) Peak to peak current modulation as a function of laser intensity.}
    \label{fig:theory_gaussian}
\end{figure}

In Fig.\,\ref{fig:theory_gaussian} we evaluate Eq.\ref{eq:Theory_Gaussian} for parameters similar to the LCLS-II CuS line: $R_{56}=0.55$\,mm, $C=5.2$, $\gamma_0=215\,\text{MeV}$, $\sigma_{\eta}=20\,\text{keV}/\gamma_0$ for a variety of laser heater amplitudes, all with field fwhm 1\,ps (i.e. power fwhm $1/\sqrt{2}$\,ps). We compare the analytical result to simulations in Elegant\,\cite{borland_elegant_2000}. At low intensity both relations are linear, while at higher intensity the current modulation begins to saturate and more terms are needed to make Eq.\,\ref{eq:Theory_Gaussian} converge. In this example, the model is suitably converged using 20 terms (for $I_s<1.4$\,GW/cm$^2$), and the remaining discrepancy between model and simulation can be attributed to the fact that the elegant lattice includes a realistic beamline with RF curvature and nonlinear optics.

%===========================================================================================
%===========================================================================================
%===========================================================================================
%===========================================================================================

\subsection{Pulse stacking}
\label{sec:analytical_pulse_stacking}
A quasi-periodic pulse train can be built by stacking consecutive Gaussian laser heater pulses with some time delay $\Delta t$. Lasing with such a current profile will create persistent temporal and spectral features that can be manipulated for use with correlation imaging techniques\,\cite{driver_attosecond_2020,ratner_pump-probe_2019}. We build a simple description showing how control of a laser pulse stacker translates to control over the beam bunching factor.

% Gaussians
 When the separation between pulses is large compared to their extent ($\Delta t\gg c\sigma_t$), we can simply superimpose copies of the current modulation in Eq.\,\ref{eq:Theory_Gaussian}. This covers most practical arrangements of a pulse stacker, since closely spaced Gaussians will overheat and spoil the entire beam.  For an infinite stack of identical Gaussian pulses separated in time by $\Delta t=2\pi/k_0 \gg c\sigma_t$ we find the bunching factor at frequency $m k_0$ to be:

\begin{align}
\label{eq:bunching_gaussian_train}
\begin{split}
b_m=\sum_{n=0}^\infty \biggr[& k_0\left(m k_0 R_{56} A_0 \right)^{2n} \frac{c \sigma_t}{C \sqrt{n}} \frac{(-1)^n}{2^{2n} (n!)^2} \times  \\
&\left(e^{-\frac{1}{2}(m k_0)^2 \tilde{\sigma}_n^2}\right) \biggr]
\end{split}
\end{align}

%interpret equation
To order $O(A_0^2)$ and for $\sigma_\eta\approx 0$, $b_m$ is maximized when the spacing between Gaussian peaks is $\Delta t =2\pi m c\sigma_t/\sqrt{3}$. For $b_1$ this roughly corresponds to placing the current maxima in Fig.\,\ref{fig:theory_gaussian}(a) halfway between the successive minima, while the harmonics are maximized by spacing the pulses farther apart. At this maximum $|b_1|\approx0.12(R_{56}A_0/\sigma_t)^2$, and $b_1$ is the largest of the harmonics. 

\begin{figure}[ht]
    \centering
    \includegraphics[width=0.45\textwidth]{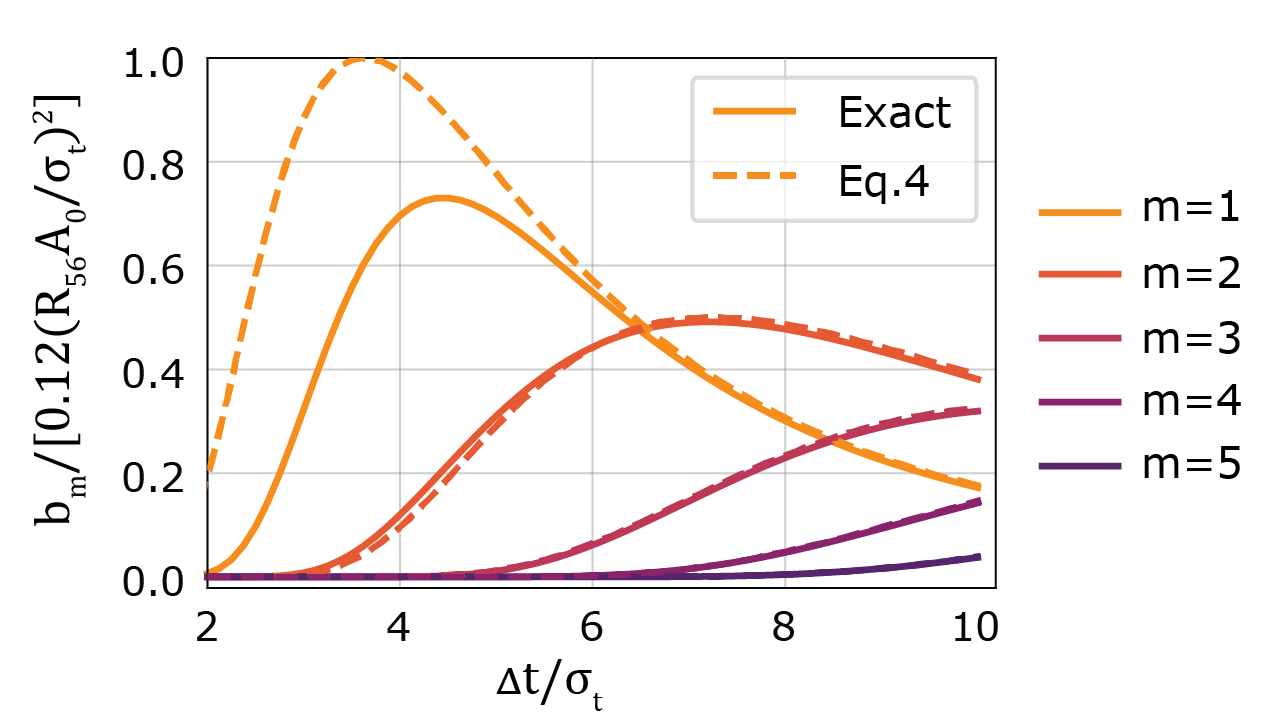}
    \caption{Magnitude of the bunching factor for a train of Gaussian laser heater pulses as a function of pulse spacing. Dashed lines are from Eq.\,\ref{eq:bunching_gaussian_train}, while solid lines are an exact calculation. When the pulses are spaced close together Eq.\,\ref{eq:bunching_gaussian_train} fails to take into account the coupling between pulses.}
    \label{fig:theory_pulse_train_gaussian}
\end{figure}

%Figure
We can see this behavior in Fig.\,\ref{fig:theory_pulse_train_gaussian}, where we have plotted the bunching factor for a train of Gaussian pulses as a function of spacing. All pulses have $\sigma_\eta=0,C=1$, and we use $R_{56}A_0/\sigma_t<<1$ such that  Eq.\,\ref{eq:bunching_gaussian_train} is well represented by $n=1$. By comparing Eq.\,\ref{eq:bunching_gaussian_train} to an exact solution, we can see that at small spacing Eq.\,\ref{eq:bunching_gaussian_train} overestimates the bunching, because it does not consider the energy spread induced by neighboring pulses. Nonetheless, it is provides a useful way to estimate the bunching factor after the first stage of compression.

%===========================================================================================
%===========================================================================================
%===========================================================================================
%===========================================================================================

\subsection{Pulse train}
\label{sec:analytical_pulse_train}
Higher frequency, more strongly periodic current modulations can be created by the chirped pulse beating technique \cite{evain_laser-induced_2010,weling_novel_1996}. Such a beam has drawn interest in mode-locking schemes, as well as for control of sidebands and other quasi-coherent features \cite{xiang_mode-locked_2012}. 

%ideal sinusoid
A perfectly sinusoidal power profile is a particularly simple case to analyze. For a spatial wavelength $2\pi/k_0$:  $A(s) = A_0\sin(k_0s)$ we find, analogous to \cite{zhang_generation_2017}, but for an unmatched laser spot (and arcsin energy distribution), we can derive the exact result:  

$$\mathcal{F}[I](k) = I_0\sum_{m}b_{2m}\delta(k+2mk_0)$$

where the bunching factor is given by

$$b_{2m}=(J_{m}(nk_0R_{56}A_0))^2 e^{-2m^2k_0^2R_{56}^2\sigma_\eta^2}$$. 

%Interpret
As expected from Eq.\,\ref{eq:general_modulation}, the bunching depends on $A_0^2$ to lowest order. But we can also see that as $k A_0 R_{56}$ is increased, the initial bunching at $b_2$ will begin to saturate and be replaced by higher order bunching terms. 

%===========================================================================================
%===========================================================================================
%===========================================================================================
%===========================================================================================

\subsection{Staged Microbunching Evolution of Current Spikes from Heater Shaping}

%Intro:
The proceeding analysis describes how a single bunch compressor translates an energy modulation into a current modulation. But in a realistic beamline, the appearance of a current modulation will begin to drive collective forces and feedback into microbunching gain.
%Coherent modulation:
Unlike the traditional analysis of microbunching gain, here we have to consider a fully coherent modulation. That is, the filamentation of phase space following compression leads to a complicated time-energy structure which can damp (or enhance) microbunching.

%Reduced model
Despite these complications, We can gain some appreciation for the effect of staging by considering the space charge impedance  \cite{huang_suppression_2004}:
\begin{equation}
    \nonumber
    \label{eq:space_charge_impedeance}
    Z_{lsc}=\frac{i Z_0}{\pi k r_b^2}\left[1-\frac{k r_b}{\gamma}K_{1}\left(\frac{k r_b}{\gamma}\right)\right]
\end{equation}
where $K_1$ is a modified Bessel function of the first time, $Z_0$ is the impedance of free space, and $r_b$ is the radius of the electron beam. In the limit $k r_b /\gamma <<1$ we find $Z\propto k$ such that the energy deviation is proportional to the derivative of the current profile (normalized here to the Alfven current $I_A=4\pi\epsilon_0 m c^3/q_e \approx 17 kA$):
\begin{equation}
    \nonumber
    \partial_z \gamma=\frac{1}{\gamma^2}\ln{\left(\frac{4 \gamma^2}{k^2 r_b^2}\right)} \partial_s \left(\frac{I(s)}{I_A}\right)
\end{equation}
For high-frequency modulations, the resulting chirp will dominate the laser heater modulation. Consequently, areas near a local current minima become positively chirped such that they can be compressed by subsequent dispersion; while areas near a local maxima will become negatively chirped such that they can be compressed by anomalous dispersion (e.g. in a dogleg).

The microbunching gain becomes more complicated in cases where the space-charge induced energy spread is comparable to the laser heater induced energy spread. In this case, the fully coherent problem needs to be analyzed. In a real beamline, one also has to consider the role of other collective effects (namely the linac wakefield and coherent synchrotron radiation) in asymmetrically chirping the bunch. In these cases it quickly becomes expedient to pursue a fully numerical, particle-tracking based approach.

%===========================================================================================
%===========================================================================================
%===========================================================================================
%===========================================================================================

\section{Numerical study of current spikes}
\begin{figure}[ht]
\centering
\includegraphics[width=0.48\textwidth]{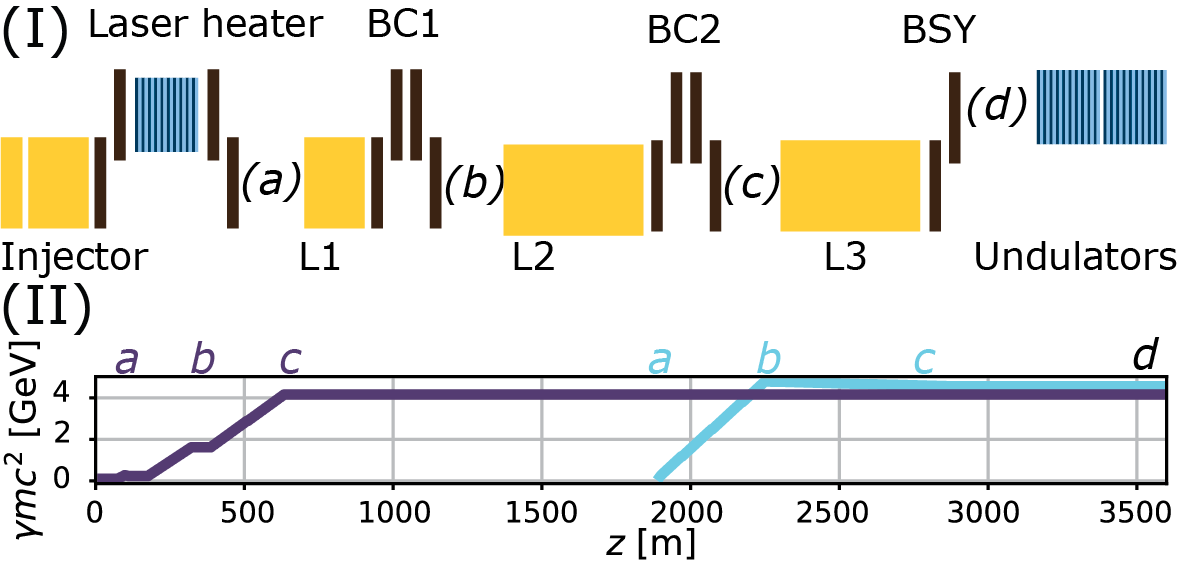}
\caption{(I) Cartoon of LCLS-II beamlines (not to scale). `L' stands for linac, 'BC' for bunch compressor, and 'BSY' for beam switch-yard. The BSY is a sequence of doglegs which are nominally dispersionless, but may be adjusted to have anomalous dispersion. (II) Central energy along the superconducting (purple) and copper (light blue) beamlines, illustrating the vast differences in space charge impedance in the two cases.}
\label{fig:beamline_diagram}
\end{figure}

%Intro to beamlines
Here we present a series of numerical studies at the LCLS-II beamlines, which are represented in a block-diagram format in Fig.\,\ref{fig:beamline_diagram}(I). LCLS-II has both a superconducting (sc) beamline and a normal conducting (copper, cu) beamline, each of which can be sent to either the soft x-ray undulators (S) or the hard x-ray undulators (H), leading to the four possibilites: scS,scH,cuH,cuS. We specialise to the two soft x-ray configurations, scS and cuS. Both beamlines consist of an injector followed by a laser heater and two bunch compression stages, before the beam switch yard routes the beam to its final destination. 

%Differences between beamlines
The high-level similarities in beamline layout belie major differences in beam dynamics. These differences are hinted at in Fig.\,\ref{fig:beamline_diagram}(II) where we show the beam energy vs z for the two beamlines. The superconducting linac accelerates at a lower gradient, has an extra 1.5\,km of transport, and a 6x larger net compression compared to the copper case. This is partially compensated for by operating at a lower peak current, but the end result is that the superconducting beam experiences larger space charge impedance and larger microbunching gain than the copper beam. Consequently, despite the fact that both beamlines have very similar BC1s, we will use a weaker laser heater modulation on the superconducting line.

%Simulation setup
We will explore how the laser-heater shapes the beam dynamics in these two cases by studying the output of start-to-end simulations at the locations marked a,b,c,d in Fig.\,\ref{fig:beamline_diagram}. Our simulations are performed in Elegant\,\cite{borland_elegant_2000}, based on the LCLS-II lattices\,\cite{woodley_lcls_nodate} using beams from dedicated injector simulations. This method has been successfully bench-marked at LCLS\,\cite{wang_benchmark_2015,qiang_start--end_2017}, and includes models for linac wakefields, longitudinal space charge, and coherent synchrotron radiation in addition to standard (nonlinear) optical maps for the static beamline elements.

%===========================================================================================
%===========================================================================================
%===========================================================================================
%===========================================================================================

\subsection{Single spike}

An isolated current spike can be used to generate attosecond x-ray pulses by the chirp-taper technique\,\cite{duris_tunable_2020}. The current spikes we show here could be used to lase directly, or they could first be used to drive CSR in a wiggler in direct analogy to \cite{duris_tunable_2020}. In either case, the goal is to create a high contrast, few-femtosecond current spike. 
%be further enhanced by anomalous dispersion in the BSY (see Fig.\,\ref{fig:beamline_diagram}) 

A single Gaussian laser heater, like that from section \ref{sec:analytical_Gaussian}, will create two current spikes after BC1. To turn this into a single, isolated, spike we shift the laser heater such that one of the current spikes forms near the head of the bunch, where the current is lower. The microbunching gain will larger for the trailing spike, which lies next to a cold, fresh beam ready to be modulated by the space-charge impedance. This allows us to form an isolated few kA spike after the second stage of bunch compression. The spike can be further enhanced by anomalous dispersion in the BSY.

\begin{figure}[]
\centering
\includegraphics[width=0.48\textwidth]{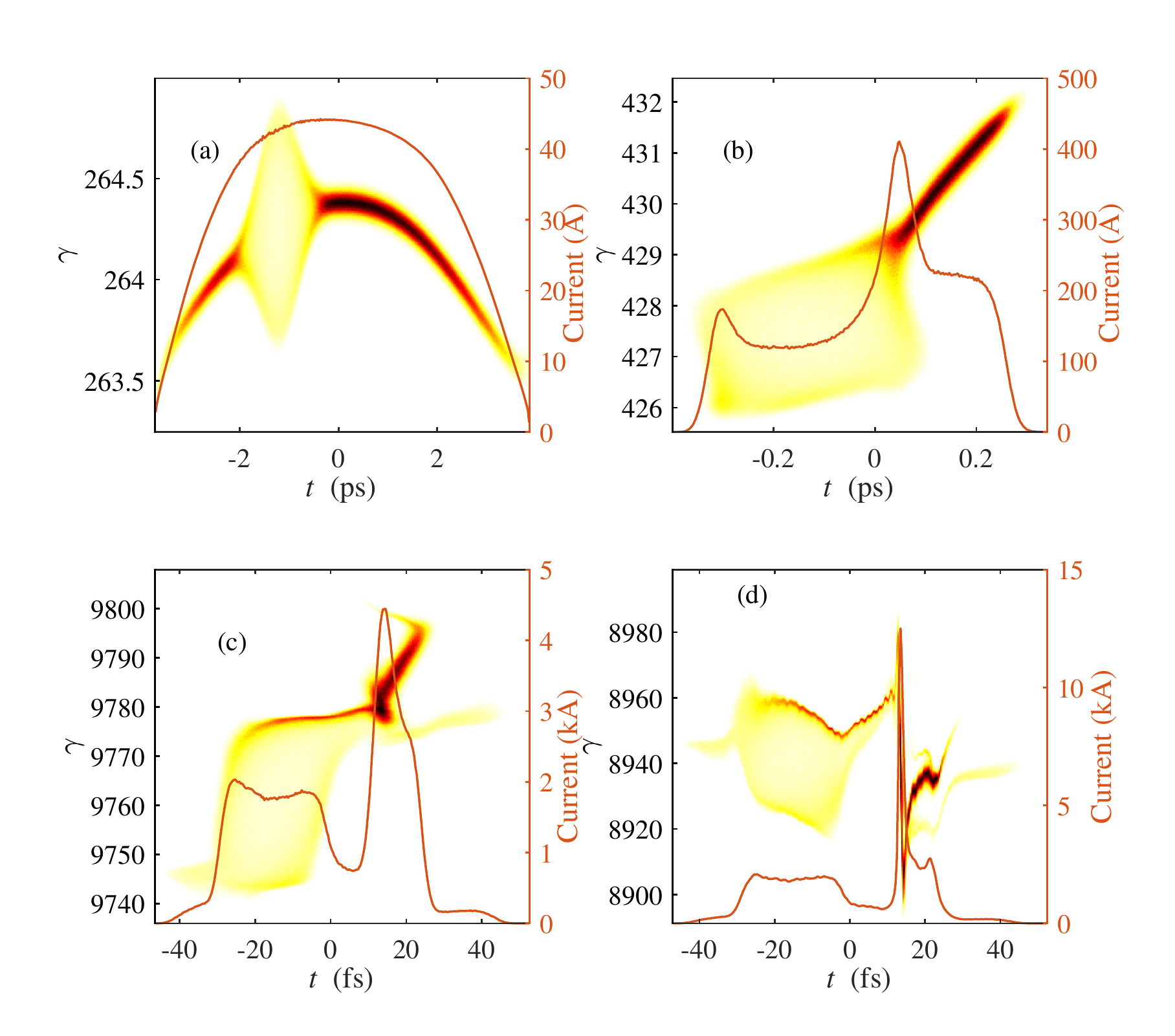}
\caption{Generation of a single high-current spike at the LCLS-II cuS beamline. Longitudinal phase space: (a) after laser heater; (b) after the first bunch compressor; (c) after the second bunch compressor; (d) at the entrance of soft x-ray undulator. See Table \ref{tab:xleap_spike}.} 
\label{fig:xleap_spike}
\end{figure}

In Fig.\,\ref{fig:xleap_spike} we illustrate this process through simulation of the LCLS-II cuS beamline. By using the laser parameters recorded in Table\,\ref{tab:xleap_spike} we introduce a large energy spread, visible in panel (a) as a Gaussian protrusion distinct from the uniform background heating. The modulation strength is made large enough to maximize the current spike which forms after BC1, as shown in panel (b). Between panels (b) and (c) the space charge impedance introduces a local-chirp along the bunch. This chirp is most pronounced, relative the slice energy spread, near the tail of the bunch, and so the second bunch compressor creates a single large spike in panel (c). Between (c) and (d) there is again a long transport with large space charge impedance, leading to a strongly chirped bunch which is further compressed by the anomalous dispersion of the BSY. At the entrance of the soft x-ray undulator we can get a high-current ($>$10\,kA) and short ($\sim$1.5\,fs) spike in the beam.

%===========================================================================================
%===========================================================================================

\begin{figure}[]
\centering
\includegraphics[width=0.48\textwidth]{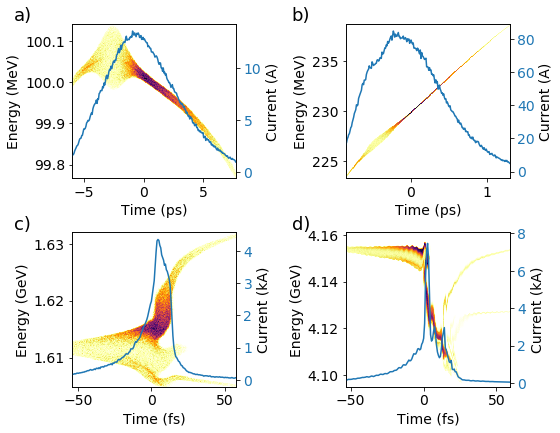}
\caption{Generation of a single high-current spike at the LCLS-II scS beamline. Longitudinal phase space: (a) after laser heater; (b) after the first bunch compressor; (c) after the second bunch compressor; (d) at the entrance of soft x-ray undulator. See Table \ref{tab:sc_xleap_spike}.}
\label{fig:SC_xleap_spike}
\end{figure}

Next we can consider a similar approach applied to the LCLS-II scS line (Fig.\,\ref{fig:SC_xleap_spike} and Table\,\ref{tab:sc_xleap_spike}). Here we use a wider laser heater pulse, which creates a much smaller modulation after BC1 (panel b) than we saw in the cuS beamline. This is appropriate because the sc beamline has lower energy and larger space-charge impedance between BC1 and BC2. As a result, this small modulation is enough to create an isolated current spike near the tail of the bunch after the large compression factor in the BC2 chicane (c). This beam can still be successfully transported through the bypass line and the BSY to the SXR undulators.

%===========================================================================================
%===========================================================================================
%===========================================================================================
%===========================================================================================
\subsection{Pulse stacking}

Here we consider creating a quasi-periodic bunch train by stacking four Gaussian pulses on top of the regular LH pulse, as described in section \ref{sec:analytical_pulse_stacking}. Such a pulse stacking arrangement can be created, for example, by sending a single Gaussian laser pulse of (power) FWHM 0.47~ps into a double interferometer that replicates the original Gaussian laser pulse into four copies. We can then tune the relative strength of each pulse with a polarizer and a waveplate, as well as the separation between the pulses by changing a time delay. This flexibility turns out to be necessary in order to compensate long-range wakefield effects and balance the compression of the resulting current spikes. 

%Parameter table
\begin{table}[] 
\begin{tabular}{|l|c|}
\hline
Beamline                     & cuS                    \\ \hline \hline
Number of stacked Gaussians  & 1                          \\ \hline
Pulse width $\tau$           & 0.95 ps                      \\ \hline  
Pulse power $P$              & 60 MW                     \\ \hline
%Min power $P$              & 60 MW                     \\ \hline
Laser waist $w_0$            & 600 $\upmu$m         \\ \hline 
Electron size (rms) $\sigma_r$     & 140 $\upmu$m         \\ \hline
Peak energy spread $A_0$      & 275 keV                \\ \hline
Uniform energy spread     & 20 keV                    \\ \hline \hline
BC1 $R_{56}$                 & -45.7 mm        \\ \hline
BC1 Compression factor $C_1$ & 5.2             \\ \hline
BC2 $R_{56}$                 & -28.6 mm         \\ \hline
BC2 Compression factor $C_2$ & 10.5              \\ \hline
\end{tabular}
\caption{Laser heater shaping parameters to generate a single high-current spike at the LCLS-II cuS beamline.}
\label{tab:xleap_spike}
\end{table}

%Parameter table
\begin{table}[h] 
\begin{tabular}{|l|c|c|}
\hline
Beamline                     & scS           \\ \hline \hline
Number of stacked Gaussians  & 1                \\ \hline
Pulse width $\tau$           & 1.5 ps          \\ \hline 
Peak power $P$              & 1.44 MW      \\ \hline
%Min power $P$           & 40 kW           \\ \hline
Laser waist $w_0$            & 180 $\upmu$m     \\ \hline
Electron size (rms) $\sigma_r$ & 130 $\upmu$m         \\ \hline
Peak energy spread $A_0$           & 70 keV      \\ \hline
Uniform energy spread     & 12 keV           \\ \hline \hline
BC1 $R_{56}$                 &  -53 mm        \\ \hline
BC1 Compression factor $C_1$ &  6.3             \\ \hline
BC2 $R_{56}$                 & -40.4 mm         \\ \hline
BC2 Compression factor $C_2$ & 51              \\ \hline
\end{tabular}
\caption{Laser heater shaping parameters to generate a single high-current spike at the LCLS-II scS beamline.}
\label{tab:sc_xleap_spike}
\end{table}

\begin{figure}[]
\centering
\includegraphics[width=0.49\textwidth]{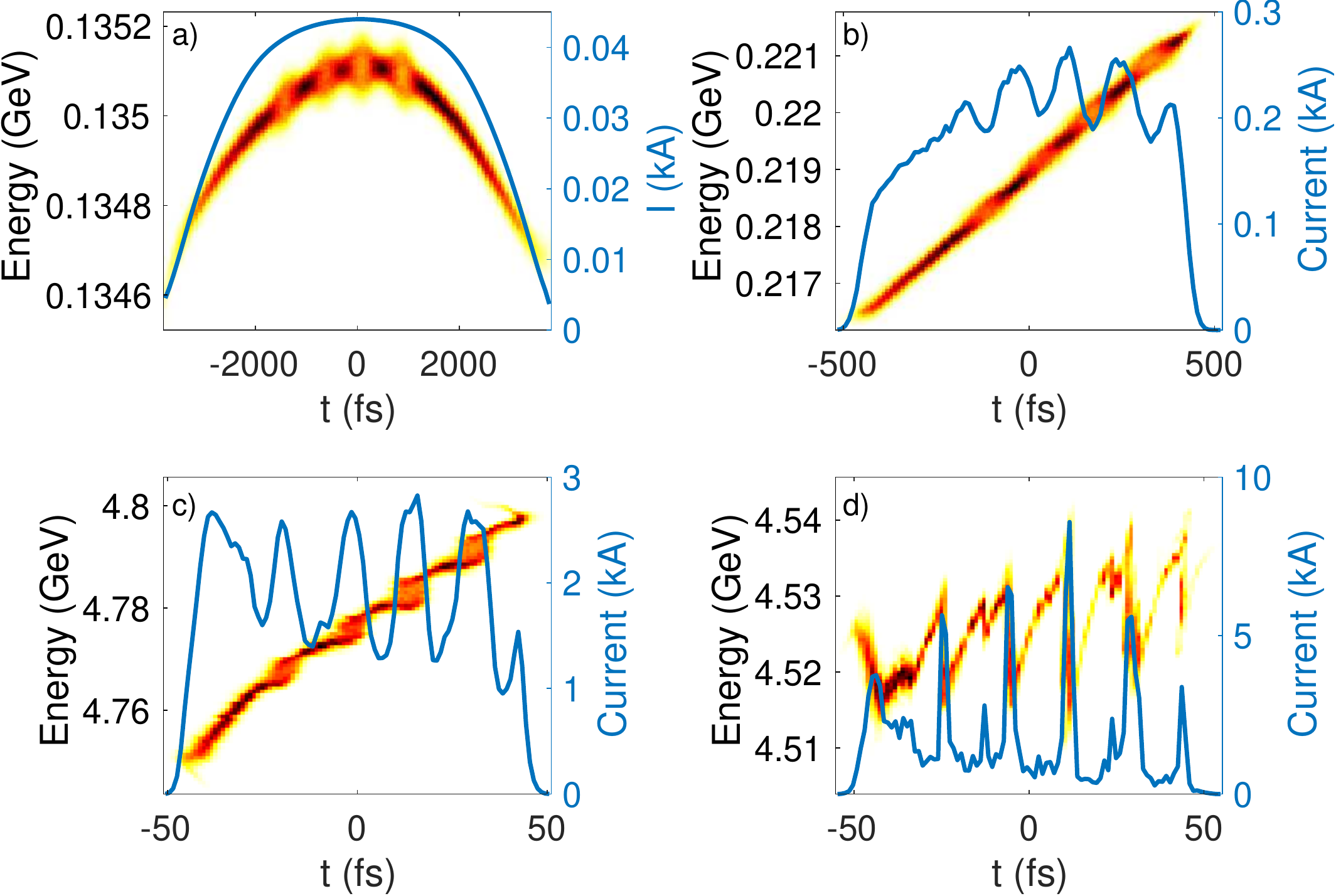}
\caption{Generation of a short bunch train at the LCLS-II cuS beamline. Longitudinal phase space: a): after the laser heater. b): after BC1. c): after BC2. d) in the middle of soft x-ray self seeding chicane, with an effective $R_{56}$ of 0.3~mm. See Table\,\ref{tab:short_bunch_train}.}
\label{fig:pulsestack_ele}
\end{figure}

In Figure\,\ref{fig:pulsestack_ele} we show a simulation of this bunch stacker at the LCLS-II cuS beamline. Each of the 4 Gaussian pulses creates two current maxima, but in-between pulses these maxima are made to overlap such that the fundamental bunching is maximized. Thus we end up with 6 current spikes, where the first and last spike are roughly half-strength. In order to get all spikes at full compression,  we make the first two pulses 20\% weaker than the last two (see table\,\ref{tab:short_bunch_train}). Additionally we note that the current spikes shown in panel (d) are located after a self-seeding chicane, such that the current spikes only become fully compressed half-way through the undulator line. This allows the spikes to be used in a self-seeding scheme and thus to create a set of discrete spectral sidebands for use in correlation spectroscopy\,\cite{driver_attosecond_2020}.

%===========================================================================================
%===========================================================================================
%===========================================================================================
%===========================================================================================
\subsection{Pulse train}

%Parameter table
\begin{table}[]
\begin{tabular}{|l|c|c|}
\hline
Beamline                     & cuS           \\ \hline \hline
Number of stacked Gaussians  & 4                \\ \hline
Pulse width $\tau$           & 0.47 ps          \\ \hline 
% Pulse separation             & 0.75, 0.75, 0.75 ps           \\ \hline
Pulse separation             & 0.8, 0.7, 0.8 ps           \\ \hline
Peak powers $P$              & 0.8,0.8,1,1 MW      \\ \hline
%Baseline power $P$           & 330 kW           \\ \hline
Laser waist $w_0$            & 450 $\upmu$m     \\ \hline 
Electron size (rms) $\sigma_r$     & 140 $\upmu$m         \\ \hline
Peak energy spread $A_0$           & 40,40,50,50 keV      \\ \hline
Uniform energy spread       & 20 keV           \\ \hline \hline
BC1 $R_{56}$                 &  -45.7 mm        \\ \hline
BC1 Compression factor $C_1$ &  5.2             \\ \hline
BC2 $R_{56}$                 & -22.7 mm         \\ \hline
BC2 Compression factor $C_2$ & 9.4              \\ \hline
\end{tabular}
\caption{Laser heater shaping parameters to generate a short bunch train at the LCLS-II cuS beamline.}
\label{tab:short_bunch_train}
\end{table}

Here we consider creating a high frequency, periodic current modulation using an intensity modulated pulse on top of the regular LH pulse, as described in section \ref{sec:analytical_pulse_train}. The intensity modulation can be generated using the chirp pulse beating technique, \cite{evain_laser-induced_2010,weling_novel_1996}.  Here a single temporally stretched Gaussian laser pulse with a linear chirp is duplicated in an interferometer.  Introducing a variable time delay to one pulse and overlapping both temporally produces a constant frequency difference proportional to the time delay and chirp, resulting in the desired intensity modulation.   

\begin{figure}[]
\centering
\includegraphics[width=0.49\textwidth]{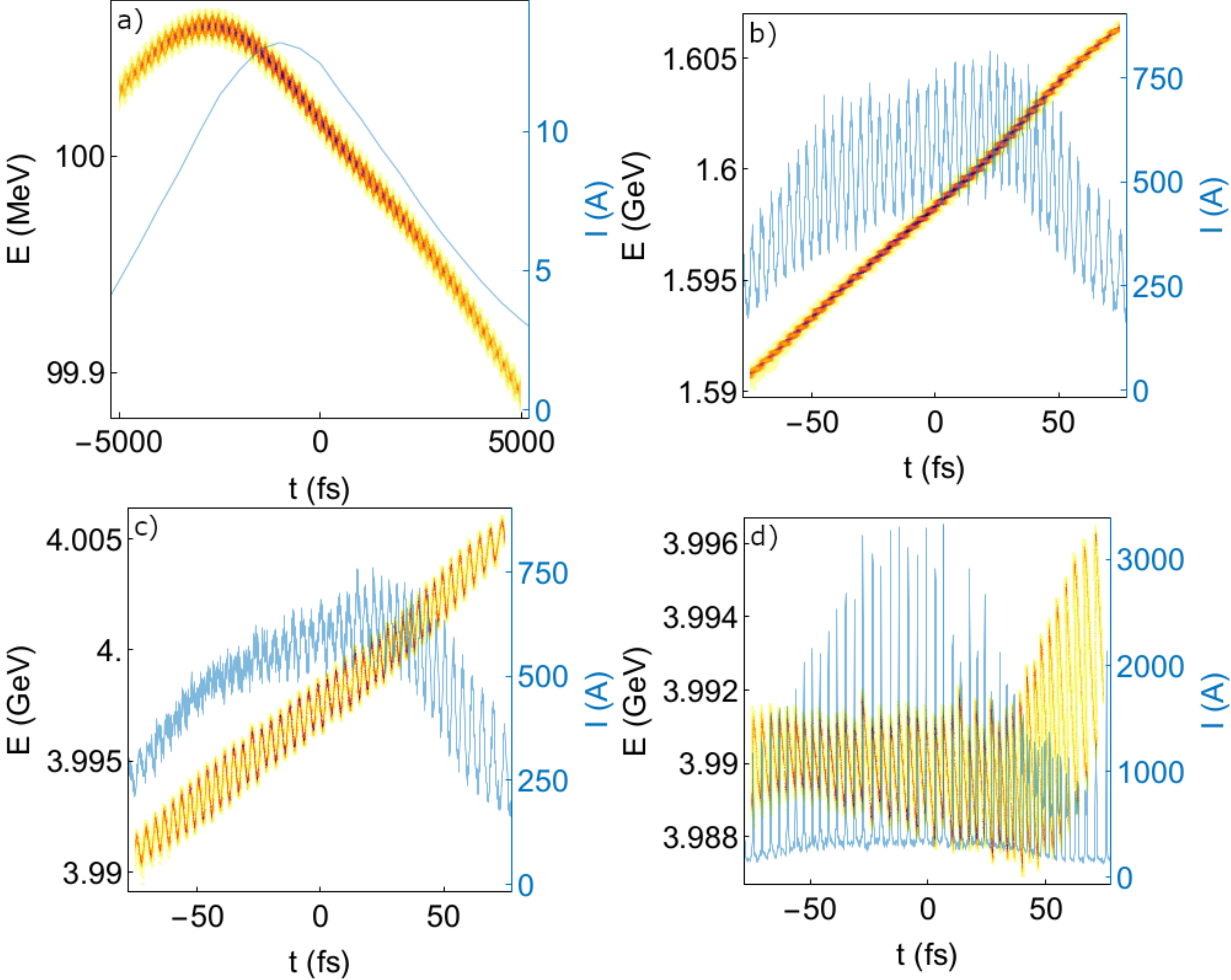}
\caption{Generation of highly periodic bunch train at the LCLS-II scS beamline. Longitudinal phase space: a) after the laser heater b) after BC2 c) after the dogleg and d) at the center of the soft x-ray undulator entrance. See Table\,\ref{tab:sc_pulse_train}.}
\label{fig:beatwave_ele}
\end{figure}

In Figure\,\ref{fig:beatwave_ele}, we show a simulation of this current modulation at the LCLS-II scS beamline.  The simulated 50 $\mu$m intensity modulation assumes a laser pulse with 10 nm fourier limited bandwidth, stretched to 20 ps with a 19.85 ps delay between split pulses, Figure\,\ref{fig:beatwave_ele}(a).  Additional simulation parameters are shown in table\,\ref{tab:short_bunch_train}.  Here the peak energy spread introduced in the laser heater is equal to the nominal heating for suppressing the microbunching instability in nominal LCLS-II operations. This energy modulation is small compared to the previous cases and will not generate significant density modulation until after BC2, producing an approximately sinusoidal current modulation with a compressed period of 1.2 $\upmu$m, Figure\,\ref{fig:beatwave_ele}(b). As the beam traverses L3, this density modulation leads to a sinusoidal energy modulation driven by longitudinal space charge (LSC).  This energy modulation can be controlled by adjusting the BC2 compression factor and/or increasing the base line LH pulse.  Furthermore, in order to produce an approximately uniform, periodic modulation across the beam, we consider the use of octupoles to reduce non-linear compression as was proposed in \cite{sudar_octupole_2020}.%PhysRevAccelBeams.23.112802

In order to reduce LSC energy modulation in the long bypass line, anomalous dispersion is introduced by the dogleg located immediately downstream of L3.  Here the total $R_{56}$ can be tuned to minimize the current modulation by adjusting compensating chicanes at the dogleg entrance and exit, Figure\,\ref{fig:beatwave_ele}(c).  Finally, this energy modulation is converted to density modulation by the self-seeding chicane located in the middle of the SXR line, Figure\,\ref{fig:beatwave_ele}(d). This allows the spikes to be used in self-seeding schemes as described in the previous section.

%Parameter table
\begin{table}[] 
\begin{tabular}{|l|c|c|}
\hline
Beamline                     & scS           \\ \hline \hline
Beatwave modulation period &  50 $\upmu$m \\ \hline
Final modulation period & 1.2 $\upmu$m \\ \hline
Peak power $P$              & 0.48 MW      \\ \hline
%Baseline power $P$           & 20 kW           \\ \hline
Laser waist $w_0$            & 2600 $\upmu$m     \\ \hline 
Electron size (rms) $\sigma_r$     & 130 $\upmu$m         \\ \hline
Peak energy spread $A_0$         & 6 keV      \\ \hline
Uniform energy spread     & 2 keV           \\ \hline \hline
BC1 $R_{56}$                 &  -55 mm        \\ \hline
BC1 Compression factor $C_1$ &  2.75             \\ \hline
BC2 $R_{56}$                 & -70.5 mm         \\ \hline
BC2 Compression factor $C_2$ & 50.8          \\ \hline
\end{tabular}
\caption{Laser heater shaping parameters to generate a highly periodic bunch train at the LCLS-II scS beamline. See Fig.\,\ref{fig:beatwave_ele}.}
\label{tab:sc_pulse_train}
\end{table}

%===========================================================================================
%===========================================================================================
%===========================================================================================
%===========================================================================================
\section{Horn suppression}

% What are horns, why we don't want them?
During compression of a long electron bunch, nonlinear optics generate caustics, typically in the form of two current spikes at the head and tail of the bunch \cite{charles_caustic-based_2016}. These current spikes, colloquially known as ``horns'', can induce CSR kicks to the beam and significantly increase its projected transverse emittance \cite{bane_measurements_2009}. 

%Static approaches to horn suppression
In the LCLS, these horns limit the efficiency of standard SASE FEL lasing. To mitigate this effect, the LCLS employs collimators in a dispersive region of the first bunch compressor to remove the parts of the electron beam affected by this third order chirp \cite{ding_beam_2016}. This approach is not suitable for high average power machines such as the LCLS-II superconducting linac or the European XFEL, where the high beam power would damage the collimators. Consequently, one must pursue new approaches, such as using compensatory nonlinear optics\,\cite{sudar_octupole_2020}.

On the other hand, advanced short pulse FEL experimental setups utilize these current horns to modulate the electron beam \cite{macarthur_phase-stable_2019,duris_tunable_2020}. Thus it is desirable to have a horn suppression mechanism which can operate with Mhz flexibility in order to interleave beams with and without horns for downstream distribution to separate users. This can be accomplished by programmaticlly switching a shaped laser heater pulse into the beamline.

A shaped laser heater can suppress the horns by heating the edges of the bunch and decreasing the phase space density that will be mapped to the caustic. At the same time, the core of the bunch should not be over-heated, which implies that the heater pulse should be shaped concave upwards, like a parabola. The new energy spread will, in-turn, create its own current spikes: from Eq.\,\ref{eq:simple_current_modulation} we can predict that a parabolic laser heater will generate a parabolic current perturbation: $\partial_s^2 (s^2)^2=12 s^2$. Design of an effective horn suppression requires balancing these two effects.

\begin{figure}[th]
    \centering
    \includegraphics[width=0.45\textwidth]{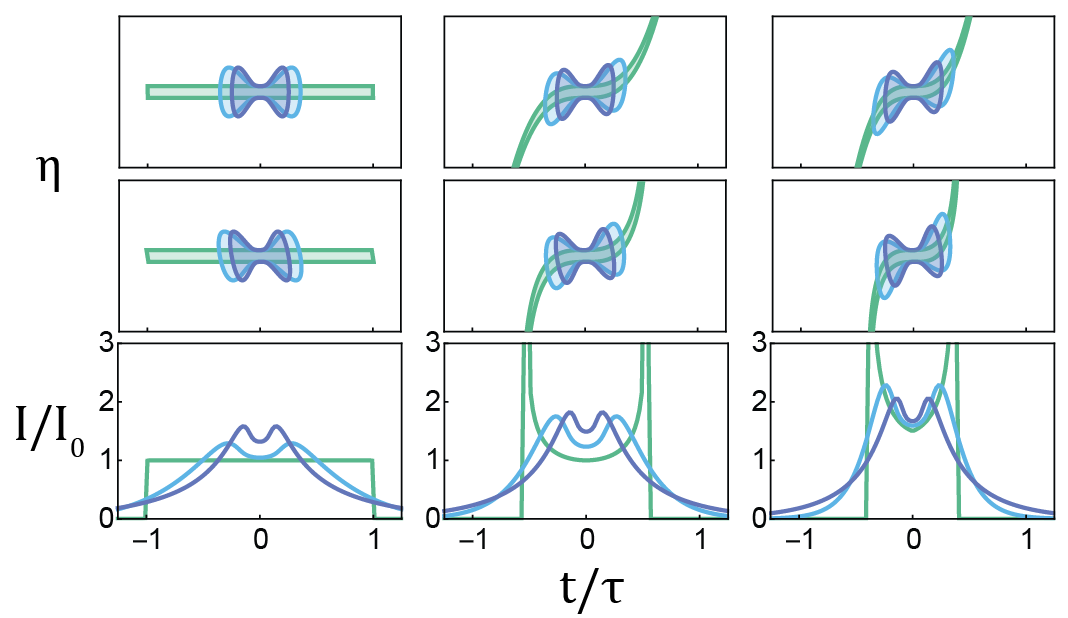}
    \caption{Illustration of heater horn suppression. The top row shows contours of longitudinal phase space before compression, the second row shows the shearing of those contours after compression, and the bottom row shows the resulting current. Each plot contains 3 different laser amplitudes of parabolic heater, and each column contains a different amplitude for the cubic chirp.}
    \label{fig:theory_horn_supression}
\end{figure}

%figure discussion
We illustrate the trade-off in Fig.\,\ref{fig:theory_horn_supression} for a simple case in which the caustic is created by a cubic chirp and suppressed by a parabolic laser heater: $\eta_f=\eta_i+\alpha t^3/\tau^3 + f_0(A_0 (t/\tau)^2)$. We also include a small incoherent energy spread (0.1\%) to speed numerical convergence. The cubic chirp increases from left to right, while the same three laser heater amplitudes are used in each plot. The amplitudes are chosen so that $\alpha =A_0$ for one curve in each plot (green, light blue, dark blue for the left, middle, and right columns respectively). When the laser heater is turned off (green curves) the caustic leads to large, narrow current spikes with lots of high frequency content. The laser heater effectively suppresses these horns and replaces them with a new perturbation. As expected from Eq.\,\ref{eq:simple_current_modulation}, the laser heater induced modulation is parabolic on-axis, while off-axis the current gradually falls to zero (recall Eq.\,\ref{eq:simple_current_modulation} applied to a current profile that was infinite in extent, while here we should include the current profile as $\partial_s^2 \left(I(s)A(s))^2\right)$).

% What about non-parabolic shape? Can we preserve more of the middle of the pulse?
Compared to the caustic, the laser heater modulation has lower peak-current and reduced high frequency content. But the parabolic profile makes for a gradual transition between the cold and hot parts of the beam, and the suppression would make a better time-filter if we used a steeper laser heater profile (such as as step function). However, a steeper laser heater profile would also create a steeper current perturbation. Beyond this qualitative reasoning, it will be necessary to use full numerical simulations. In fact, in practical cases the current modulations from either the caustic or the laser heater are small after BC1, but can still influence the nonlinear chirp going into BC2. %[rework this idea? Move it to a new section?] 

%===========================================================================================
%===========================================================================================
%===========================================================================================
%===========================================================================================
\subsection{Simulations}

As we did for the current spike generation, we will use elegant\,\cite{borland_elegant_2000} to model the beam dynamics associated with horn suppression at the LCLS-II cuS and scS beamlines. In place of a perfectly parabolic laser heater we use a more practical arrangement of two Gaussian heater pulses. We also include a direct comparison to the case with the Gaussian heater pulses, in order to emphasize the difference between the caustics and heater-induced current modulation. 

\begin{figure}[th]
    \centering
    \includegraphics[width=0.45\textwidth]{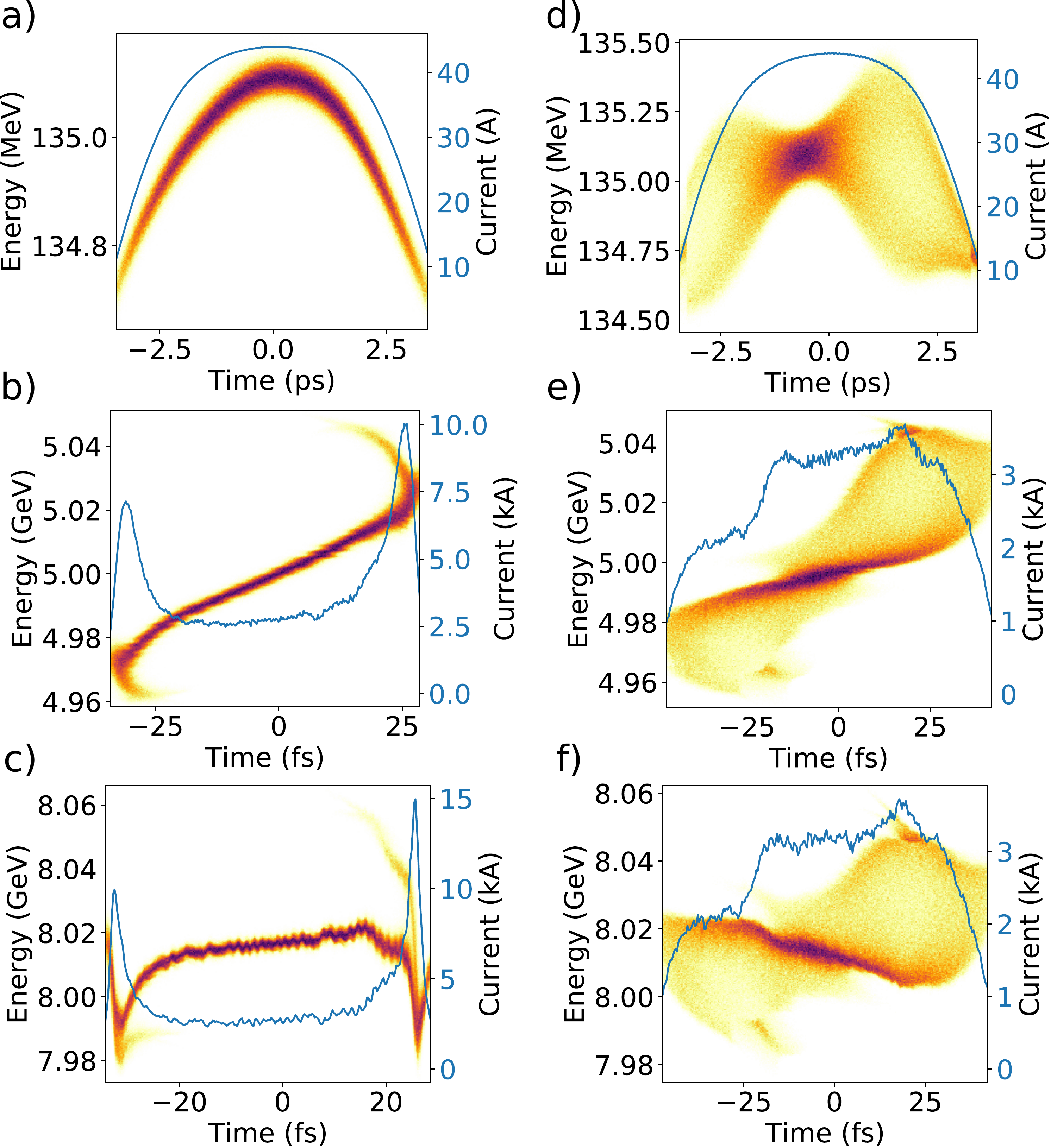}
    \caption{Elegant simulations a-c) without and d-f) with laser heater horn collimation for the LCLS-II normal conducting, copper linac. Figures a) and d) show the longitudinal phase space after the laser heater with and without the pairs of Gaussian pulses. Figures b) and e) show the beam after the second bunch compressor. Figures c) and f) show the beam at the entrance of the hard x-ray undulators. See Table\,\ref{tab:horn_supression}.}
    \label{fig:lcls horns}
\end{figure}

Figure~\ref{fig:lcls horns} shows simulations of the LCLS Cu beamline for a 250~pC bunch charge compressed to 3~kA. 
Horns formed within the second bunch compressor increase the projected horizontal emittance from 0.50~$\mathrm{\mu}$m to 1.38~$\mathrm{\mu}$m, and a subsequent CSR kick from the LCLS second dog leg leads to a final emittance of 2.78~$\mathrm{\mu}$m at the undulator line. Applying two Gaussian laser pulses each of duration 1~ps and delayed by 4~ps, heats the head by 190~keV and the tail by 220~keV, dispersing the charge which would have otherwise formed the horns in the bunch compressor. This mitigates some of the CSR kick-induced emittance growth while producing a similar peak current. The resulting projected emittance after the second bunch compressor and undulator entrance is a more modest 0.70~$\mathrm{\mu}$m.

%Parameter table
\begin{table}[bh]
\begin{tabular}{|l|c|c|}
\hline
Beamline                     & LCLS     & LCLS-II scS               \\ \hline \hline
Number of stacked Gaussians  & 2        & 2                  \\ \hline
Pulse width $\tau$           & 1 ps     & 1.1 ps                 \\ \hline  
Pulse separation             & 5.5 ps   & 7 ps                    \\ \hline
Peak powers $P$              & 2, 2.75 MW   & 200, 200 kW                  \\ \hline
%Baseline power $P$              & 500 kW   & 50 kW                  \\ \hline
Laser waist $w_0$            & 280$\upmu$m & 180$\upmu$m        \\ \hline 
Electron beam size $\sigma_x$            & 140$\upmu$m & 115$\upmu$m        \\ \hline 
Peak energy spread $A_0$       & 120,180 keV &   33, 33 keV             \\ \hline
Uniform energy spread     & 20 keV  & 8 keV                     \\ \hline \hline
BC1 $R_{56}$                 &  -48.0 mm & -52.5 mm                           \\ \hline
BC1 Compression factor $C_1$ &   5.2 & 5.8                            \\ \hline
BC2 $R_{56}$                 &  -24.9 mm & -40.3 mm                           \\ \hline
BC2 Compression factor $C_2$ &  15.2 & 17.2                             \\ \hline
\end{tabular}
\caption{Laser heater shaping parameters to suppress caustic formation at the LCLS and at the LCLS-II scS beamlines.}
\label{tab:horn_supression}
\end{table}

% check length of bypass line.
The LCLS-II superconducting linac is positioned 2~km upstream of the copper linac, but shares the same FEL beamlines with the copper linac, necessitating the long transport line shown in Fig.\,\ref{fig:beamline_diagram}(II). Current horns formed in the bunch compressor of the superconducting linac can induce strong wakefields along this long transport line, increasing the energy spread of the electron beam and degrading FEL lasing performance. 

\begin{figure}[h]
    \centering
    \includegraphics[width=0.45\textwidth]{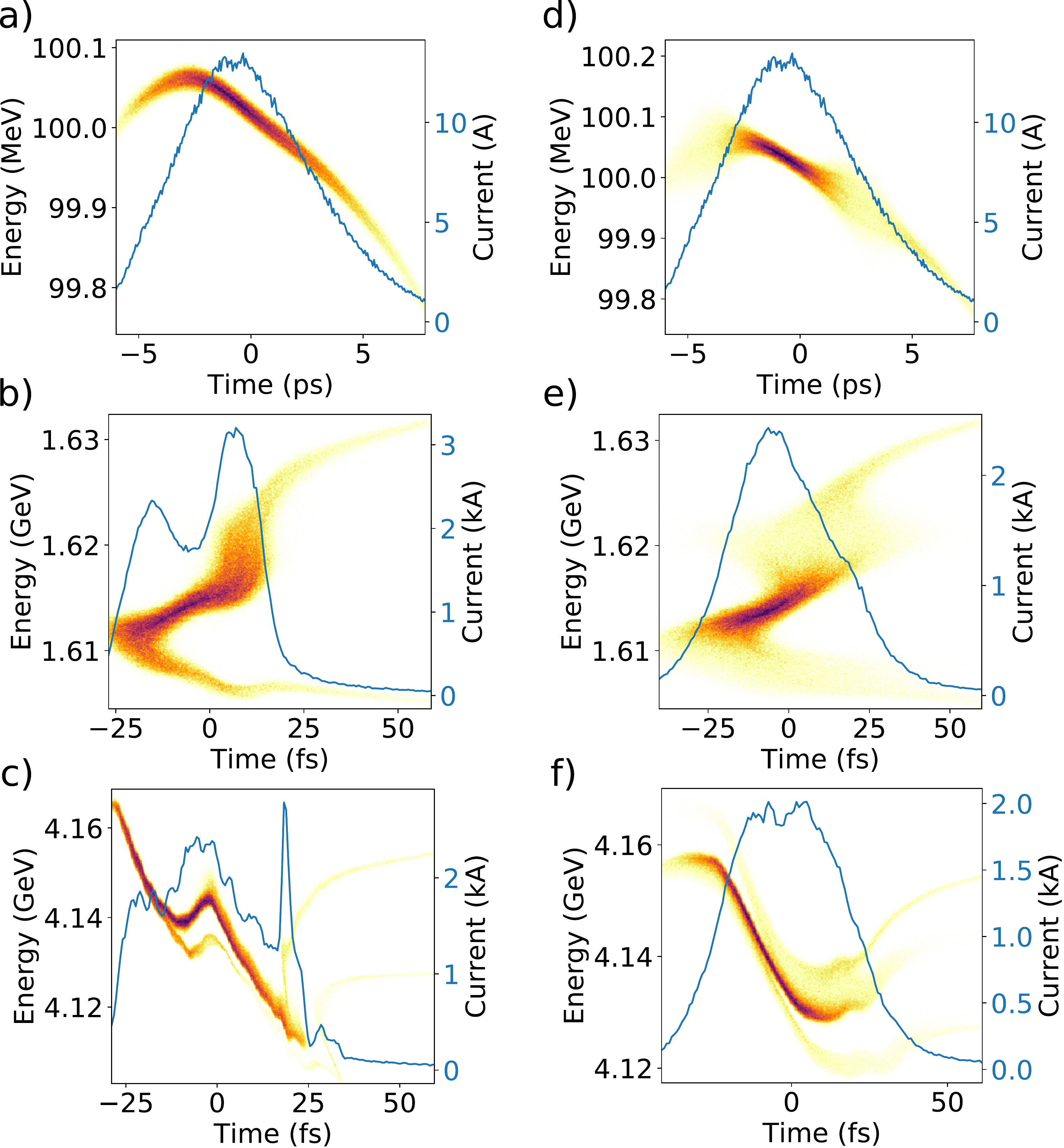}
    \caption{Elegant simulations a-c) without and d-f) with laser heater horn collimation for the LCLS-II superconducting linac. Figures a) and d) show the longitudinal phase space after the laser heater with and without the pairs of Gaussian pulses. Figures b) and e) show the beam after the second bunch compressor. Figures c) and f) show the beam at the entrance of the soft x-ray undulators. See Table\,\ref{tab:horn_supression}.}
    \label{fig:lcls2 horns}
\end{figure}

Simulations of the LCLS-II scS beamline are shown in Fig.~\ref{fig:lcls2 horns} using the baseline scenario of a 100\,pC beam formed by a temporally flat-top cathode laser. Increasing the baseline compression from 800~A to 2~kA results in high current horns and significant energy variation along the beam. Heating the head and tail of the beam with 1.1~ps pulses spaced 7~ps apart results in a beam of similar duration (50~fs FWHM) and current but with 50\% less energy variation. Furthermore, several additional horizontal and vertical bends are needed to transport the superconducting beam to the soft x-ray FEL line, each of which can spoil the emittance via CSR. The horizontal and vertical projected emittances at the undulator line are reduced from 2.68~$\mathrm{\mu}$m and 0.66~$\mathrm{\mu}$m without laser heater shaping to 1.87~$\mathrm{\mu}$m and 0.50~$\mathrm{\mu}$m with.

\section{Conclusions}
In this article we have explored the versatility of laser heater shaping as a technique to manipulate the beam current. Our findings demonstrate that a time dependent laser heater profile is capable of shaping the bunch current to best suit a variety of operational modes. Thanks to the flexibility in tailoring the repetition rate and macropulse structure, we forsee the ability interleave multiple operational modes for multiple beamlines fed by the same linac at high repetition rate. For example, in a high repeition rate XFEL operating multiple undulators simultaneously, one could apply different shaping modes to the bunches going to each beamline.

We have used simple analysis followed by start-to-end simulations to give a broad understanding of the capabilities of this technology. The simple kinetic analysis we present is of interest for giving physical intuition to the process by which laser heater shaping acts on the beam. It shows that, for small modulations, the modulation amplitude scales linearly with the laser power and has a shape given by the second derivative of the power profile. For larger modulations, the current valleys begin to saturate and the peaks steepen. The simplified model we use is helpful for optimizing simulations and will be instructive when building online controls to help integrate laser heater shaping into an operational context.

Our start-to-end simulations illustrate how laser-heater shaping could be used to optimize the beam at LCLS-II. Each case study we discuss optimizes the beam for a unique purpose: to create current spikes for attosecond lasing,  to make a bunch train, and finally to suppress caustics and reduce CSR induced emittance growth. Furthermore, we have shown that these techniques can work across a variety of accelerator beamlines at the LCLS-II facility, indicating the robustness of the technique. Our results provide a basis for understanding the wide scope of laser-heater shaping, and thus pave the way for XFEL facilities to simultaneously optimize performance for distribution to multiple users.

\begin{acknowledgments}
The authors would like to thank Zhirong Huang for useful discussions and suggestions.
This work was supported by the Department of Energy, Office of Science under Contracts No. DE-AC02-76SF00515, and the Office of Basic Energy Science Accelerator and Detector Research Program.
\end{acknowledgments}

\bibliography{references} 

\end{document}